\documentclass[12pt]{article}
\usepackage[utf8]{inputenc}
\usepackage{newtxmath} 

\usepackage[a4paper,width=170mm,top=15mm,bottom=20mm]{geometry}
\DeclareMathOperator*{\argmin}{arg\,min}
\newtheorem{question}{Question}
\usepackage{amsfonts,amssymb}

\usepackage{hyperref}
\usepackage{xcolor}
\hypersetup{
    colorlinks,
    linkcolor={red!50!black},
    citecolor={blue!50!black},
    urlcolor={blue!80!black}
}
\usepackage[authoryear]{natbib}
\usepackage{float}
\usepackage{authblk}
\usepackage{orcidlink}
\usepackage{setspace}
\usepackage{subfig}

\title{
Bayesian Stability Selection and Inference on Selection Probabilities}

\author[1]{Mahdi Nouraie\orcidlink{0000-0002-4792-4994}}
\author[1]{Connor Smith\orcidlink{0000-0002-3955-4348}}
\author[1,2]{Samuel Muller\orcidlink{0000-0002-3087-8127}\thanks{Address for Correspondence: samuel.muller@mq.edu.au}}
\affil[1]{School of Mathematical and Physical Sciences, Macquarie University}
\affil[2]{School of Mathematics and Statistics, The University of Sydney}

\date{}
\begin{document}
\begin{spacing}{1}
\maketitle
\begin{abstract}
\noindent{Stability selection is a versatile framework for structure estimation and variable selection in high-dimensional setting, primarily grounded in frequentist principles. In this paper, we propose an enhanced methodology that integrates Bayesian analysis to refine the inference of selection probabilities within the stability selection framework. Traditional approaches rely on selection frequencies for decision-making, often disregarding domain-specific knowledge. Our methodology uses prior information to derive posterior distributions of selection probabilities, thereby improving both inference and decision-making. We present a two-step process for engaging with domain experts, enabling statisticians to construct prior distributions informed by expert knowledge while allowing experts to control the weight of their input on the final results. Using posterior distributions, we offer Bayesian credible intervals to quantify uncertainty in the variable selection process. Furthermore, we demonstrate how the integration of prior knowledge reduces the variance of selection probabilities, thereby improving the stability of decision-making. Our approach preserves the versatility of stability selection and is suitable for a broad range of structure estimation challenges.}
\end{abstract}
\end{spacing}
\section{Introduction}\label{s1}
The estimation of discrete structures, including graphs and clusters, and the selection of variables represent long-standing and fundamental problems in statistics \citep{meinshausen2010stability}.
Stability selection, a widely recognized resampling-based variable selection framework, determines the selection of variables based on their selection frequencies \citep{meinshausen2010stability}; which means the frequency with which variables are included in estimated models across multiple random resamples of observations. Stability selection has remained an active area of research since its inception (see, for example, \citet{bodinier2023automated} and \citet{Kissel2024}).

This is somewhat surprising, given that the discussion section of \citet{meinshausen2010stability} includes some remarks suggesting the potential to explore stability selection from a Bayesian perspective, and despite the substantial attention stability selection has received since its inception, we have found no studies in the literature that integrate it with Bayesian analysis. In this paper, we explore how the Bayesian framework can enhance stability selection by incorporating prior information, offering a comprehensive tool for decision-making and inference.

We demonstrate the methodology using variable selection in the context of linear regression. However, the applications of this approach are not limited to linear regression, as stability selection itself is not limited to this problem. 

In the influential work by \citet{meinshausen2010stability}, the stability selection framework was introduced, offering a transformative approach for identifying stable variables in structure estimation tasks. The stability selection in \citet{meinshausen2010stability} uses sub-samples of observations with half-size of data to perform variable selection and selects the variables that are consistently identified as important variables across the majority of sub-samples as stable variables. Later, \citet{shah2013variable} introduced complementary pairs stability selection, a method where selection frequencies are calculated by counting the number of times a variable is selected in models fitted to complementary 50\% sub-samples. 

\citet{meinshausen2010stability} and \citet{shah2013variable} defined selection probabilities solely based on selection frequencies, interpreting these as observed proportions. Although stability selection has shown significant utility across a diverse range of applications, as highlighted by \citet{LindleyFuture} and \citet{Jaynes2003}, the foundation of probability lies within the domain of human knowledge, while frequency is rooted in the external world. We define the selection probability as the probability that a variable is relevant to the response variable through a fixed model, taking into account both data and background knowledge. Our objective is to use selection frequencies to adjust prior probabilities. In addition, we examine a non-informative prior for selection probabilities, using it as a proxy for frequentist selection probabilities. We then compare the variance of posterior selection probabilities with and without prior knowledge, demonstrating how the incorporation of prior knowledge reduces the variance, thereby enhancing the stability of decision-making.

In this context, we introduce a novel yet intuitive approach for integrating domain expert knowledge into the selection process when such expertise is available, while explicitly managing the inherent subjectivity of the final outcomes. This method facilitates effective communication with data owners by providing transparency regarding how their expertise is incorporated into the analysis and the extent of its influence on the results.

The structure of this paper is organized as follows. Section \ref{s2} details the proposed methodology. Section \ref{s3} describes the real and synthetic datasets used to apply the methodology. The results of the application of the method to real and synthetic data are presented in Section \ref{s4}. Section \ref{s5} offers some remarks on Bayesian stability selection, and finally, Section \ref{s6} provides a summary and concludes the paper.

\section{Methodology}\label{s2}

In this section, we introduce a methodology that integrates Bayesian analysis into the stability selection framework. Although we illustrate the proposed method in the context of the stability selection version of \citet{meinshausen2010stability}, the proposed methodology is adaptable to other versions of stability selection, for example, as presented in \citet{shah2013variable} and \citet{beinrucker2016extensions}. 

We consider  a dataset $\mathcal{D} = \{(\boldsymbol{x}^\top_{i}, y_{i})\}_{i=1}^{n}$ comprising of a univariate response variable $y_{i} \in \mathbb{R}$ and a $p$-dimensional vector of fixed covariates $\boldsymbol{x}^\top_{i} \in \mathbb{R}^{p}$. We assume that the pairs $(\boldsymbol{x}^\top_{i}, y_{i})$ are independent and identically distributed (i.i.d.).

The linear regression model is formally defined as $   \boldsymbol{Y} = \boldsymbol{
\beta_0} + X\boldsymbol{\beta} + \boldsymbol{\varepsilon},$
where $X \in \operatorname{mat}(n \times p)$ denotes the design matrix, $\boldsymbol{Y} \in \mathbb{R}^{n}$ denotes the $n$-dimensional vector representing the univariate response variable, $\boldsymbol{\beta_0} \in \mathbb{R}^{n}$ denotes the constant $n$-dimensional vector of intercept term, $\boldsymbol{\beta} \in \mathbb{R}^{p}$ denotes the $p$-dimensional vector of regression coefficients for the $p$ non-constant covariates, and $\boldsymbol{\varepsilon} \in \mathbb{R}^{n}$  denotes the $n$-dimensional vector of random errors. 

It is assumed that $E(\boldsymbol{Y}\mid X)$ is linear in parameters, $E(\boldsymbol{\varepsilon}\mid X) = 0$, $\text{Var}(\boldsymbol{\varepsilon} \mid X) = \sigma^2 < \infty$, and $\text{Cov}(\boldsymbol{\varepsilon}, X_{j}) = 0$, for $j \in \{1,2,\dots,p\}$, where $X_{j}$ denotes the $j$th column of $X$. It is also assumed that the columns of $X$ are not linearly dependent of each other, and the components of $\boldsymbol{\varepsilon}$ are independent of each other.

\subsection{Stability Selection}

\sloppy
In variable selection, following the terminology of \citet{meinshausen2010stability}, it is generally assumed that the covariates can be divided into two distinct groups: the signal group $S := \{k \neq 0 \mid \beta_k \neq 0\}$ and the noise group $N := \{k \neq 0 \mid \beta_k = 0\}$, where $S \cap N = \emptyset$. A primary goal of a variable selection algorithms is to accurately estimate the signal group $S$.

The stability selection framework requires an appropriate selection algorithm to be implemented within it. One widely used method for variable selection in linear regression setting is the Least Absolute Shrinkage and Selection Operator \citep[Lasso;][]{tibshirani1996regression}, a regularization technique that employs the $\ell_1$-norm to perform shrinkage on vanishing coefficients. The Lasso estimator is defined as
\begin{equation}\label{eqn:Lasso}
\boldsymbol{\hat{\beta}_0}(\lambda),\boldsymbol{\hat{\beta}}(\lambda) = \argmin_{\boldsymbol{\beta_0} \in \mathbb{R}^{n}, \boldsymbol{\beta} \in \mathbb{R}^{p}} \left(\|\boldsymbol{Y} - \boldsymbol{\beta_0} - X\boldsymbol{\beta}\|_{2}^{2} + \lambda \sum_{k = 1}^{p} |\beta_k|\right),
\end{equation}
where $\lambda \in \mathbb{R}^{+}$ is the Lasso regularization parameter. The set of non-zero coefficients $\hat{S}(\lambda) := \{k \neq 0 \mid \hat{\beta}_{k}(\lambda) \neq 0\}$ can be identified using the solution from Equation~\eqref{eqn:Lasso} via convex optimization.

\sloppy
\citet{meinshausen2010stability} proposed applying the selection algorithm (here, Lasso) on the different random sub-samples of observations and defined the stable set
$\hat{S}^{\text{stable}} := \{j \mid \max_{\lambda \in \Lambda} (f_{j}^{\lambda}) \geq \pi_{\text{thr}}\};\; j = 1, \ldots, p$,
where $\Lambda$ denotes the set of regularization values, $0 < \pi_{\text{thr}} < 1$ denotes a threshold for decision-making in variable selection, and $f_{j}^{\lambda}$ denotes the selection frequency of the $j$th variable given the regularization parameter $\lambda$.

\subsection{Bayesian Stability Selection}
In this paper, we conceptualize stability selection as a repetitive experiment. Given that one of the primary objectives of this paper is to perform inference on selection probabilities, it is essential to carefully examine both the systematic and random components of this experiment. To ensure that the variability in the results of stability selection across different iterations arises solely from changes in the training dataset, and without loss of generality, we restrict the hyper-parameter set $\Lambda$ to a single value, denoted as $\lambda$, which is consistently applied across all iterations of the stability selection procedure. 

The value $\lambda$ is selected by identifying the largest value of $\lambda$ such that the prediction error remains within one standard error of the minimum error obtained from the cross-validation \citep{hastie2009elementss} on all the data $\mathcal{D}$. If exploration of different $\lambda$ values is desired, one may employ a grid of $\lambda$ values and repeat the process for each one. 

During each iteration of the stability selection procedure, a random sub-sample comprising half the size of the original dataset $\mathcal{D}$ is used. The Lasso model is fitted to the current sub-sample, and the binary selection results are recorded as a row in the binary selection matrix $M(\lambda) \in \operatorname{mat} (B \times p)$, where $B$ denotes the total number of sub-samples. In this context, $M(\lambda)_{bj} = 1$ indicates that the $j$th variable is selected as a member of the signal set $\hat{S}(\lambda)$ by applying Lasso to the $b$th sub-sample. In contrast, $M(\lambda)_{bj} = 0$ signifies that the $j$th variable is included in the noise set $\hat{N}(\lambda)$ when Lasso is applied to the $b$th sub-sample. 

By assuming independence in the selection of variables, similar to \citet{bodinier2023automated}, we consider a Bernoulli distribution for the selection of each variable in stability selection iterations
$M(\lambda)_{bj} \stackrel{\text{ind}} \sim \text{Bernoulli}(\Pi_j^{\lambda});\; j = 1, \ldots, p$,
where $\Pi_j^{\lambda}$ denotes the selection probability of the $j$th variable given $\lambda$ that is, $P(j \in S(\lambda) \mid \lambda, K)$ where $K$ denotes the expert's prior knowledge about the relevance of the covariate. 

Therefore, the total number of iterations in which the $j$th variable is selected during stability selection, assuming independence of sub-samples, follows a Binomial distribution, that is
\begin{equation}\label{eqn:Binom}
    n_{j}:= \sum_{b = 1}^{B}M(\lambda)_{bj} \sim \text{Binomial}(B, \Pi_j^{\lambda});\quad j=1, \ldots,p.
\end{equation}

To go further, it is necessary to select an appropriate prior distribution for $\Pi_j^{\lambda}$. The Beta distribution $\text{Beta}(\alpha, \beta)$ has been of great interest in modeling selection probabilities  (e.g., see \citet{kohn2001nonparametric,  bottolo2010evolutionary, liang2018bayesian,  
staerk2024metropolized}) and has support $[0,1]$. Moreover, when the shape parameters of the Beta distribution are greater than or equal to $1$, that is, $\alpha,\beta \geq 1$, they are easily interpretable, facilitating the incorporation of background knowledge by directly translating it into a statistical object. 

In literature, several approaches have been proposed to incorporate prior knowledge into the prior distributions of selection probabilities when using the Beta distribution. \citet{kohn2001nonparametric} suggested fixing the expectation and variance of the prior distribution based on available background knowledge and determining the parameters of the distribution by solving two simultaneous linear equations. \citet{hu2009bayesian} proposed the prior distribution $\text{Beta}(\delta p, (1 - \delta)p)$, where $0 < \delta < 1$ is termed the precision parameter. This approach is designed to maintain the influence of the prior in high-dimensional settings and discourage large model sizes. \citet{ley2009effect} proposed fixing $\alpha = 1$ and determining $\beta$ by solving the corresponding equations after specifying the expectation and variance of the distribution based on the background knowledge. \citet{castillo2012needles} demonstrated that the use of small $\alpha$ and large $\beta$ is more effective in high-dimensional settings, where the number of covariates significantly exceeds the number of observations. They recommended setting $\alpha = 1$ and $\beta = p$, to promote sparsity in the model. Lastly, \citet{liang2018bayesian} recommended the use of empirical Bayes methods, as described in \citet{Carlinemp1996}, to determine the parameters of the Beta distribution. 

In the Beta distribution, the shape parameters $\alpha$ and $\beta$ offer significant insight into the form and characteristics of the distribution. When $\alpha,\beta > 1$, the relative magnitudes of $\alpha$ and $\beta$ determine the skewness of the distribution. Specifically, if $\alpha > \beta$, the corresponding density is left skewed, with the mode shifted toward $1$. In contrast, if $\beta > \alpha$, the density is right skewed, with the mode shifted toward $0$. When $\alpha = \beta$, the Beta distribution is symmetric. Importantly, the Beta distribution encompasses the standard Uniform distribution $\mathcal{U}(0,1)$ as a special case when $\alpha = \beta = 1$, which represents a non-informative flat prior over the interval $[0, 1]$. Therefore, expert knowledge can be incorporated into the Beta distribution by expressing beliefs through an according choice of the shape parameters $\alpha$ and $\beta$, thereby incorporating prior information into the distribution in a straightforward manner. We will further elaborate on this in Section \ref{S23}.

An additional advantage of the Beta distribution is its conjugacy with the Binomial distribution, which allows for the straightforward estimation of posterior distributions through reparametrization, thereby obviating the need for complex numerical methods in posterior estimation.

By employing the Beta distribution as prior distribution of selection probability for the $j$th variable we have
\begin{equation}\label{eqn:prior}
\pi(\Pi_j^{\lambda} \mid K) =   
 \frac{({\Pi_j^{\lambda}})^{\alpha_j-1}(1-\Pi_j^{\lambda})^{\beta_j-1}}{B(\alpha_j, \beta_j)}
;\quad j=1, \ldots,p\quad \text{and} \quad \alpha_j,\beta_j \geq 1,
\end{equation}
where $\alpha_j$ and $\beta_j$ are the shape parameters of the $j$th Beta prior distribution and $B(\alpha_j, \beta_j) = \int_{0}^{1} x^{\alpha_j - 1} (1 - x)^{\beta_j - 1} \, dx$.

Given $\Pi_{j}^{\lambda}$, the probability of observing the data which is referred to as the likelihood function \citep{jeffreys1934probability}  based on Equation~\eqref{eqn:Binom} is
\begin{equation}\label{eqn:like}
    \mathcal{L}(n_{j} \mid \Pi_j^{\lambda}, K) = \binom{B}{n_j} (\Pi_j^{\lambda})^{n_j} (1 - \Pi_j^{\lambda})^{B - n_j};\quad j=1, \ldots,p.
\end{equation}

The posterior distributions can be obtained using the prior distributions in the form of Equation~\eqref{eqn:prior} and the information of the rows as observed samples using Equation~\eqref{eqn:like}. In other words, each row of $M(\lambda)$ is considered an independent experiment and the prior distribution will be updated in light of the information of the rows. Using Bayes' theorem \citep{Bayestheorem}, the posterior distribution is proportional to the product of the prior distribution and the likelihood function \citep{jeffreys1934probability}, that is
\begin{equation*}
\pi(\Pi_j^{\lambda} \mid n_{j}, K)
    \propto (\Pi_j^{\lambda})^{\alpha_j - 1 + n_j} (1 - \Pi_j^{\lambda})^{\beta_j - 1 + B - n_j},
\end{equation*}
which is of the form of the Beta distribution $\Pi_j^{\lambda} \mid n_{j},K \sim \text{Beta}(\alpha'_j = \alpha_j + n_j, \beta'_j = \beta_j + B - n_j)$.

The posterior distribution allows for the evaluation of the relevance of variables based on the characteristics of the distribution. In this paper, the expectation values of the posterior distributions of selection probabilities are used as a benchmark for decision-making; however, other approaches such as the one proposed by \citet{Barbieri2004} are also applicable.

\subsection{Assigning non-informative and informative priors}\label{S23}

Here, we propose a method for translating background knowledge into the Beta distributions. As mentioned before, when $\alpha_j,\beta_j \geq 1$, an increase in $\alpha_j$ reflects a greater number of selections, thereby shifting the probability distribution towards higher values, whereas an increase in $\beta_j$ reflects a greater number of non-selections, shifting the distribution towards lower values. The sum $\gamma_j := \alpha_j + \beta_j$ corresponds to the total number of pseudo-observations considered on the basis of prior knowledge. It is essential that this total does not exceed the number of observed stability selection iterations $B$, to ensure that the prior does not disproportionately influence the results. Although Bayesian statistics does not impose restrictions on the use of priors that may dominate the data, our objective is to maintain a balanced interplay between the objectivity of the data and the subjectivity of the priors, thereby ensuring fair decision-making. As a result, if the stability selection outcomes do not support selection of a variable, the prior will not be too strong to select it and vice versa. In fact, we argue that informed variable selection requires a minimum level of support from both data-driven results and prior knowledge. In the absence of prior information, we recommend using $\alpha_j = \beta_j = 1$ as a non-informative prior distribution. 

Bayesian variable selection methods integrate background knowledge into the prior distribution by translating expert knowledge into statistical parameters. When such background knowledge is available, we propose a two-step approach to convert this knowledge into the Beta distribution. The first step involves assessing the level of subjectivity included in the final results, while the second step evaluates the perceived relevance of the variables under consideration. Specifically, we pose two questions to data owners for each variable subject to selection. 
\begin{question}\label{Qstn-1}
    Considering that the final results are a synthesis of both your opinions and data-driven insights, what percentage of the final results would you prefer to be influenced by your prior knowledge, with a maximum of $50\%$?
\end{question}
We denote the response to Question \ref{Qstn-1} by $\Tilde{\zeta}_j$. For example, $\Tilde{\zeta}_j = 50\%$ indicates that the final result incorporates both $B$ pseudo-observations and $B$ stability selection iterations, with each component contributing $50\%$ of the total weight. The value $\Tilde{\zeta}_j$ facilitates determining the number of pseudo-observations $\gamma_j$ accounted for by the background knowledge. Therefore, the value of $\Tilde{\zeta}_j$ determines the sum of the prior parameters for the Beta distribution of the $j$th variable as follows
\begin{equation}\label{eqn:Q1} 
\Tilde{\zeta}_j = \frac{\gamma_j}{\gamma_j + B}.
\end{equation}
Conceptually, $\lfloor\gamma_j\rfloor$ rows are added to $M(\lambda)_{j}$ based on the provided prior information. 

\begin{question}\label{Qstn-2}
Based on your knowledge and expertise, what percentage of the data-driven experiments do you expect to indicate that the $j$th variable is relevant?
\end{question}
We denote the response to Question \ref{Qstn-2} by $\Tilde{\xi}_j$, where the value $\Tilde{\xi}_j$ represents the perceived relevance of the $j$th variable. Specifically, $\Tilde{\xi}_j$ corresponds to the proportion of 1's in the binary pseudo-observations of the $j$th variable. Based on this, we can calculate the values of $\alpha_j$ and $\beta_j$ as follows
\begin{equation}\label{eqn:Q2} 
\alpha_j =\lfloor \Tilde{\xi}_j \gamma_j \rfloor \quad \text{and} \quad \beta_j = \gamma_{j} - \alpha_{j}.
\end{equation}
Thus, $\Tilde{\xi}_j$ determines the values of the prior parameters, providing insight into the relevance of the $j$th variable based on prior knowledge provided. We expect that $\Tilde{\xi}_{j}$ will tend to values closer to either $0\%$ or $100\%$, as it is unreasonable that prior expertise would justify assigning a variable an importance of approximately $50\%$. In the absence of prior information, the use of non-informative priors is recommended.

Therefore, Equations~\eqref{eqn:Q1} and~\eqref{eqn:Q2} offer an interpretable procedure to incorporate background knowledge into statistical distributions. Equation~\eqref{eqn:Q1} sets the total number of pseudo-observations $\gamma_j$, while Equation~\eqref{eqn:Q2} determines the value of the shape parameters $\alpha_j$ and $\beta_j$ according to the perceived relevance of variables. This approach effectively integrates expert opinions into the Beta distribution treating $B$ as a shared parameter between subjective and objective aspects of the procedure to control the degree to which prior knowledge impacts the final results.

To the best of our knowledge, existing Bayesian variable selection methods typically engage in inquiries regarding variable relevance, similar to the second question we proposed. Our approach seamlessly integrates expert's background knowledge into the selection process, thus enhancing the management of the subjectivity-objectivity trade-off.

In this paper, previously published results are used to construct informative priors; however, the lack of access to domain experts for our data examples limits our ability to directly demonstrate the effectiveness of the proposed approach. These priors serve as a proxy for expert knowledge, thereby allowing us to simulate the incorporation of expert insights into our methodology.

\begin{figure}[H]
    \centering
    \subfloat[$n_{j} = 30$]
    {\includegraphics[width=0.32\textwidth]{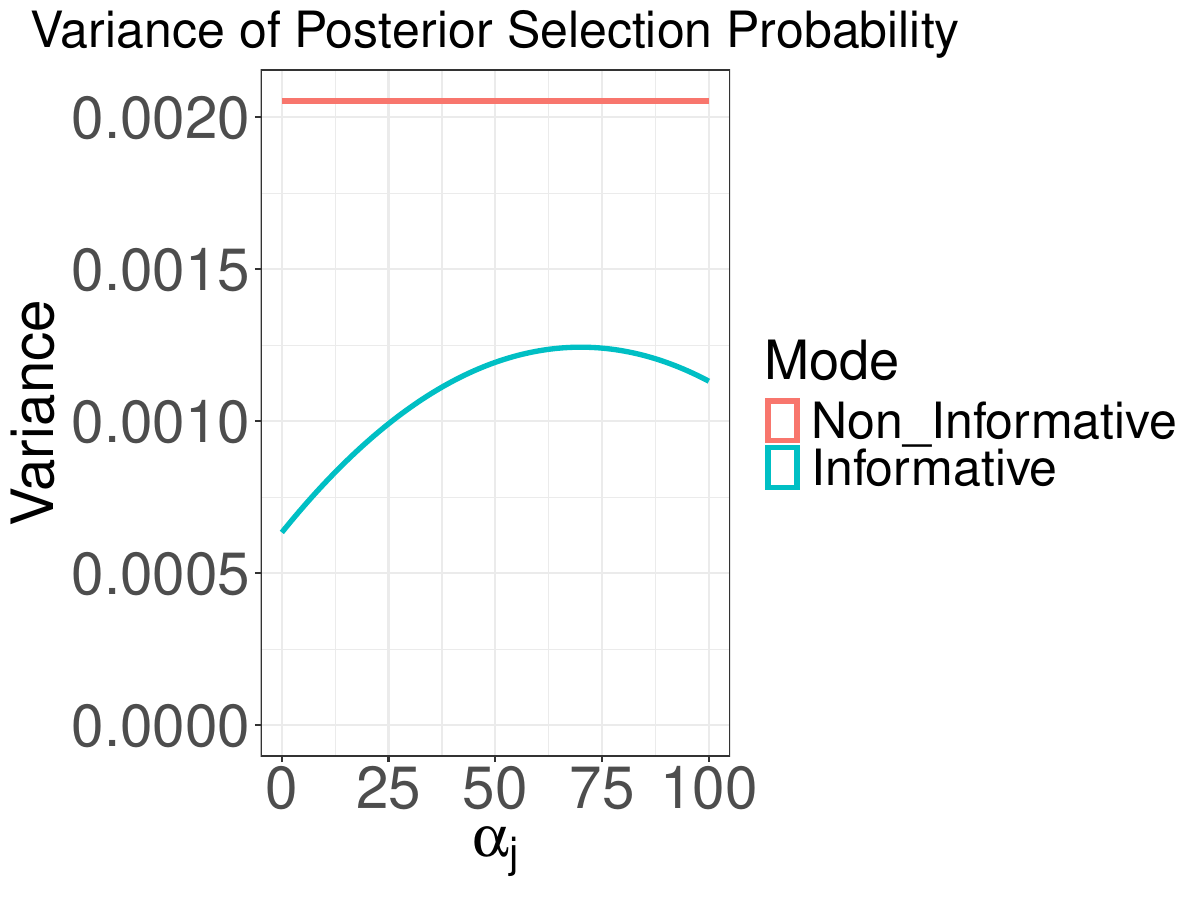}\label{fig:V1}}\hfill
    \subfloat[$n_{j} = 50$]
    {\includegraphics[width=0.32\textwidth]{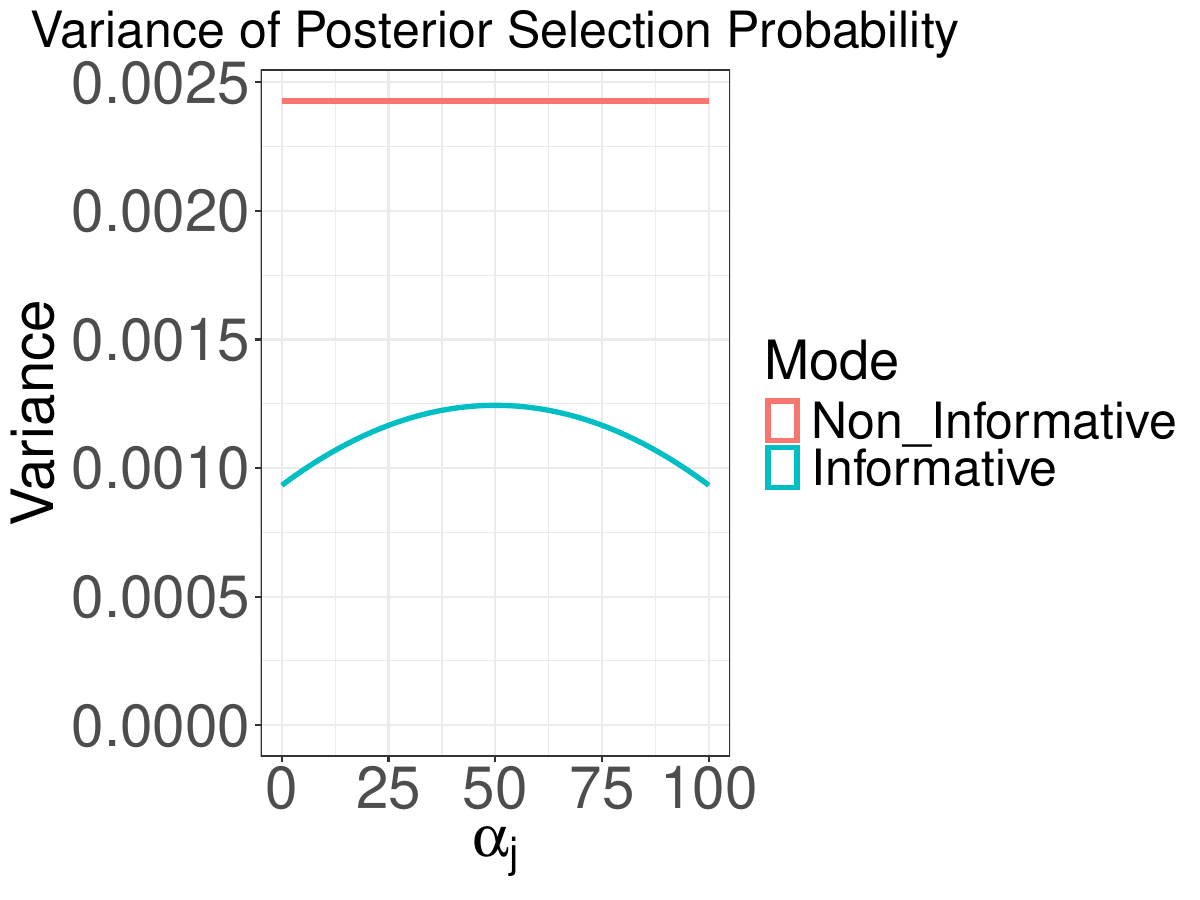}\label{fig:V2}}\hfill
    \subfloat[$n_{j} = 70$]
    {\includegraphics[width=0.32\textwidth]{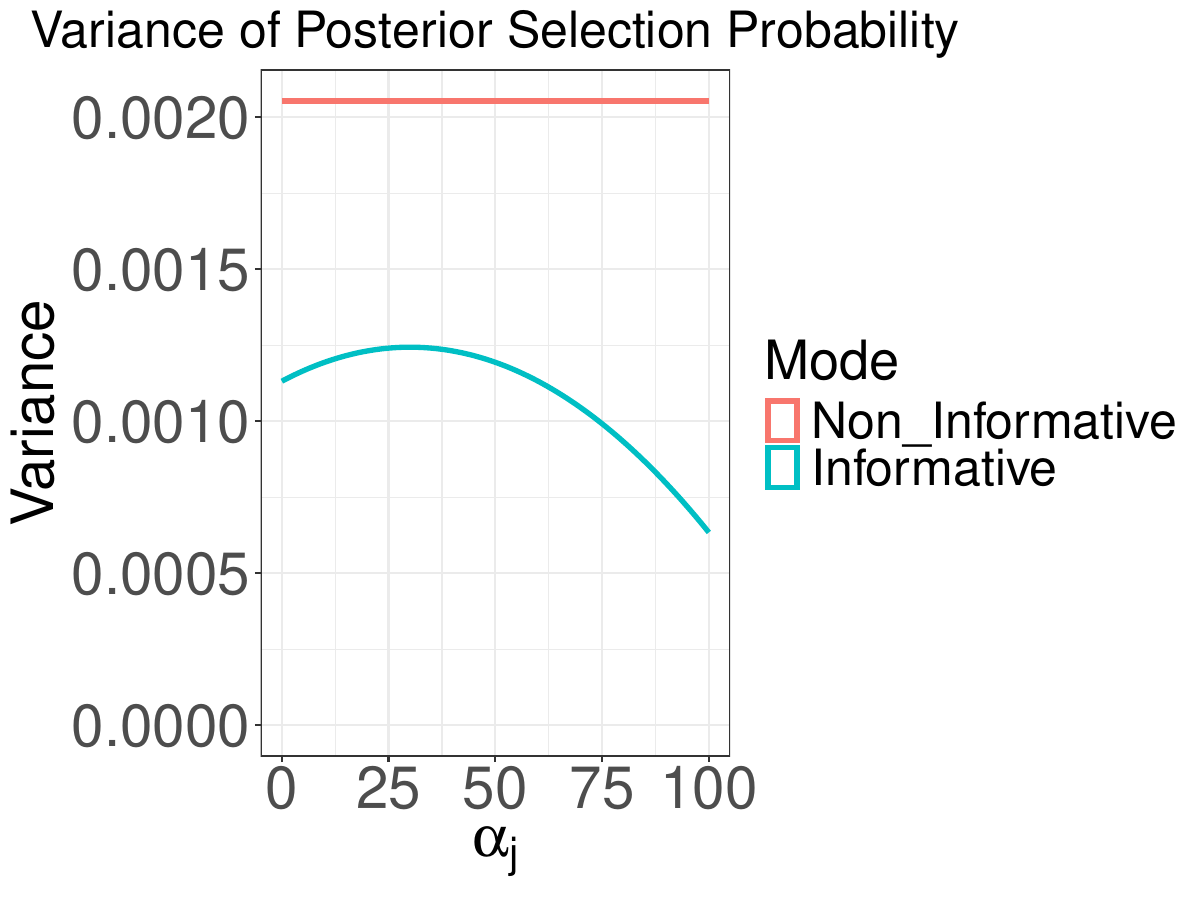}\label{fig:V3}}
    \caption{Variance of the posterior selection probability as a function of $\alpha_{j}$ when $B = 100$ and $n_{j} \in \{30, 50, 70\}$}
    \label{fig:three_variance_figures}
\end{figure}

The variance of the posterior selection probability in the non-informative mode is solely dependent on $B$ and $n_j$. However, in the informative mode, it is also influenced by the parameters $\alpha_j$ and $\beta_j$. Here, for visualization purposes, we assume $\gamma_j = B$, which implies that $\Tilde{\zeta} = 50\%$. Figure \ref{fig:three_variance_figures} illustrates the variance for both non-informative and informative modes, with the total number of sub-samples held constant while $n_{j}$ takes three distinct values. The effect of $\alpha_{j}$ is evaluated across its entire admissible domain. As shown in the figure, for all three values of $n_{j}$, the variance in the informative mode is consistently lower than that in the non-informative mode for all values of $\alpha_{j}$. According to \citet{nogueira2018stability}, stability is inversely related to variance. Therefore, based on  the Figure \ref{fig:three_variance_figures}, it can be concluded that in the scenario considered, the informative mode consistently leads to more stable decisions.

\begin{figure}[H]
    \centering
    \subfloat[$n_{j} = 10$]
    {\includegraphics[width=0.5\textwidth]{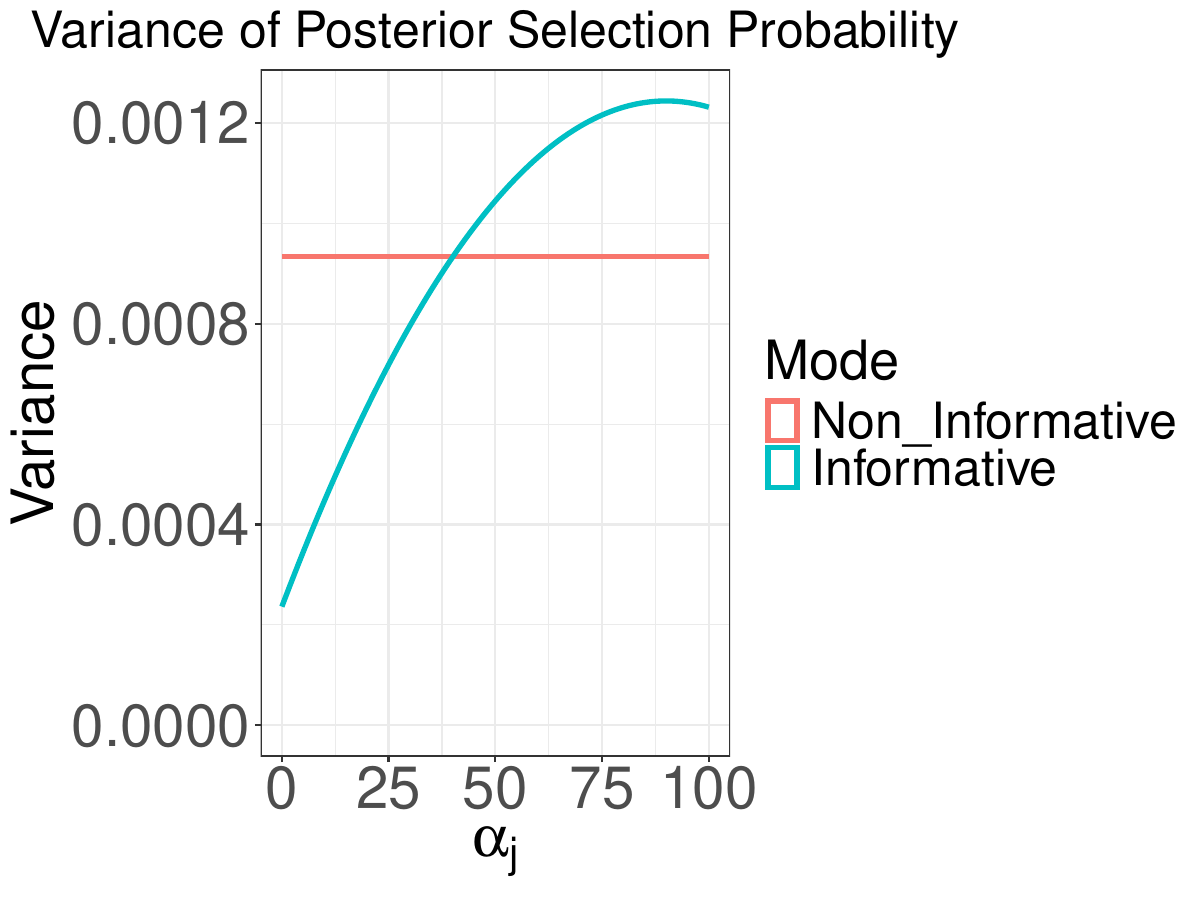}\label{fig:V12}}\hfill
    \subfloat[$n_{j} = 90$]
    {\includegraphics[width=0.5\textwidth]{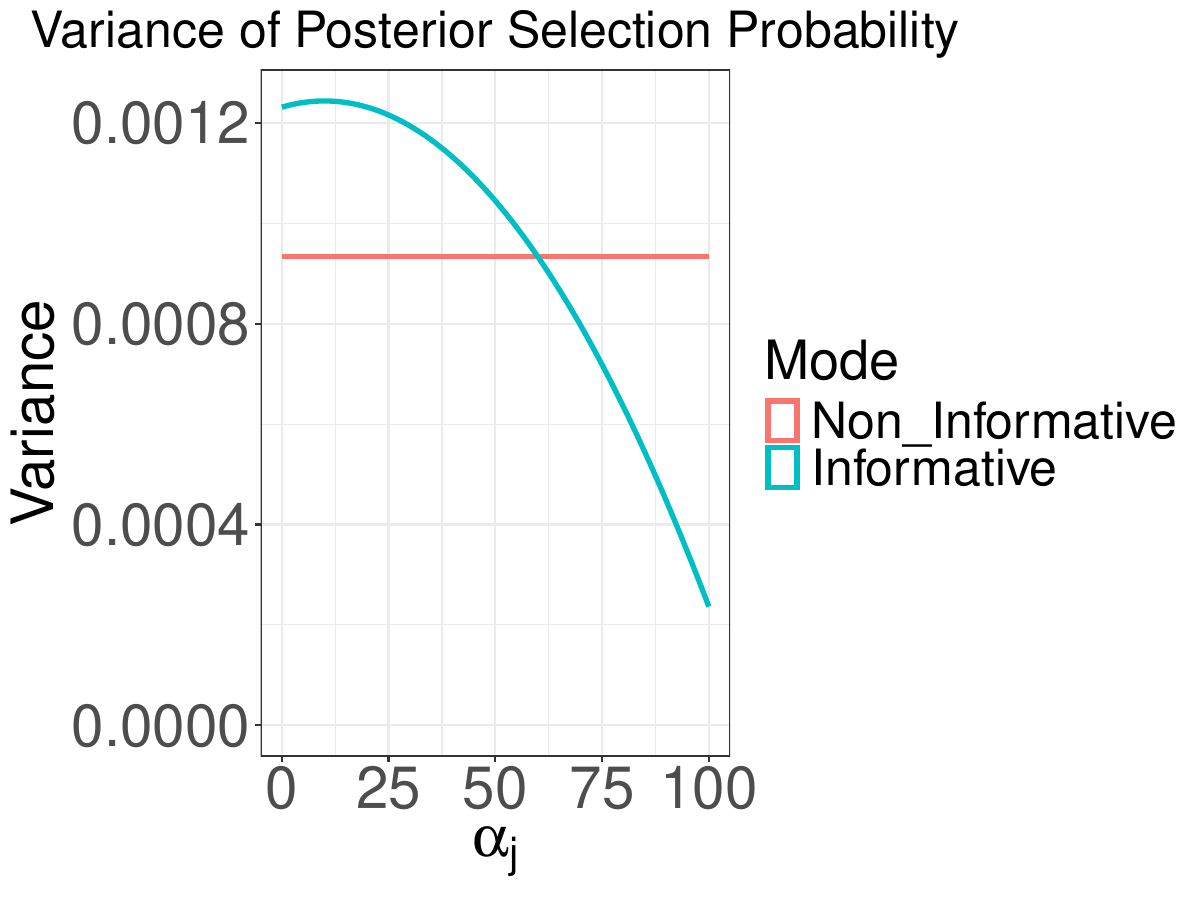}\label{fig:V22}}\hfill

    \caption{Variance of the posterior selection probability as a function of $\alpha_{j}$ when $B = 100$ and $n_{j} \in \{10, 90\}$}
    \label{fig:two_variance_figures}
\end{figure}

The dynamics changes for extreme values of $n_j$. As shown in Figure \ref{fig:two_variance_figures}, for  small or large values of $n_j$, the variance in the informative mode is lower than that in the non-informative mode only when prior knowledge aligns with the results derived from the data. In cases where there is a significant discrepancy between expert knowledge and data-driven results, the variance in the informative mode exceeds that in the non-informative mode. In such scenarios, it would be valuable to further investigate the underlying reasons for the strong disagreement between expert opinion and data.

\begin{figure}[H]
    \centering
    \subfloat[Non-informative prior]
    {\includegraphics[width=0.5\textwidth]{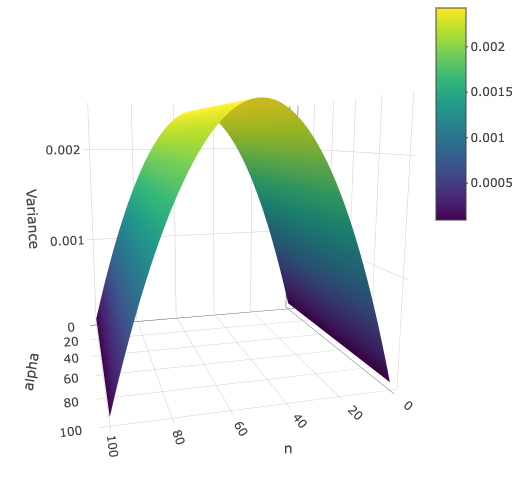}\label{fig:3D1}}\hfill
    \subfloat[Informative prior]
    {\includegraphics[width=0.5\textwidth]{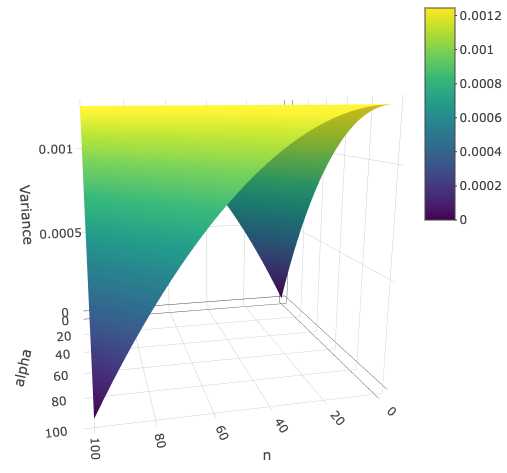}\label{fig:3D2}}\hfill

    \caption{Variance of the posterior selection probability as a function of $\alpha_{j}$ and $n_{j}$ when $B = 100$}
    \label{fig:two_3D_figures}
\end{figure}

Figure \ref{fig:two_3D_figures} illustrates the variance of the posterior selection probability by allowing both $\alpha_{j}$ and $n_{j}$ to take all admissible values. As shown in Figure \ref{fig:3D1}, in the non-informative prior case, the variance remains independent of $\alpha_{j}$, while $n_{j}$ exhibits a quadratic effect on the variance. The highest variance occurs when $n_{j}$ is at the midpoint of its range. On the other hand, Figure \ref{fig:3D2} shows that, for informative priors, maximum variance occurs when $\alpha_{j} = B - n_{j}$, indicating that prior knowledge actively opposes the selection of variables based on data-driven results. In Figure \ref{fig:3D2}, the minimum variance is observed when both $\alpha_{j}$ and $n_{j}$ are strongly aligned, either in favor or against the selection of the variable.

In Section \ref{s4}, we present the results obtained from applying the Bayesian stability selection to both synthetic and real datasets, which will be introduced in Section \ref{s3}.

\section{Datasets}\label{s3}
We evaluate Bayesian stability selection using both synthetic data and two real bioinformatics datasets, as detailed below.

\subsection{Synthetic Data}

In the first scenario, we generate data with a sample size of $n = 50$ and the number of predictor variables $p = 500$, where the predictor variables $\boldsymbol{x}^\top_{i}$ are independently drawn from $\mathcal{N}(0, \Sigma)$ where $\Sigma$
is the identity matrix except for the elements $\Sigma_{12} = \Sigma_{34} =  \Sigma_{35} = \Sigma_{45} = 0.8$ and their symmetric counterparts; therefore, two sets of covariates are correlated: $C_1 = \{x_1,x_2\}$ and $C_2 = \{x_3, x_4, x_5\}$. The response variable is linearly dependent only on the first six predictor variables with coefficient vector
$\boldsymbol{\beta} = (
  0.9, 0.9, 0.7, 0.7, 0.7, 1.5, 0, \ldots, 0)^{\top}$ and the error term $\boldsymbol{\varepsilon}$ is an i.i.d.\ sample of the zero-centered Gaussian distribution with standard deviation $\sigma = 2$. A roughly similar scenario, in which the predictor variables exhibit correlation, is considered in the discussion section of \citet{meinshausen2010stability}, by Kirk et al. They showed that in such cases, Lasso tends to select one of the variables from the correlated group at random. As a result, the selection frequencies either favor one variable with a high selection frequency or result in low selection frequencies for all variables of the group. To address this issue, they suggested the use of elastic net \citep{zou2005regularization}, which combines the Lasso $\ell_1$-penalty with the ridge regression $\ell_2$-penalty  \citep{tikhonov1963solution, hoerl1970ridge}. We employ the elastic net in this context and demonstrate that the problem in such a case persists. In fact, this is one of the situations in which incorporating background knowledge can assist the algorithm in overcoming data inherent ambiguity.

In the second scenario, we change the structure of $\Sigma$ and $\boldsymbol{\beta}$. Here, $\Sigma$ is assumed to follow a decaying structure, defined as $\Sigma_{ij} = 0.9^{|i - j|}$. The coefficient vector is specified as $\boldsymbol{\beta} = (0.5, 0.4, 0.3, 0.2, 0, \ldots, 0)^{\top}$. As in the first scenario, we illustrate how Bayesian stability selection can enhance the decision-making process regarding variable relevance when prior information is available.

\subsection{Riboflavin Data}
\sloppy
The first real dataset we use is the well-established `Riboflavin' data focused on the production of riboflavin (vitamin B2) from different Bacillus subtilis, provided by Dutch State Mines Nutritional Products, which is available through the \texttt{hdi} R package \citep{hdi-package}. The dataset comprises a single continuous response variable, representing the logarithm of the riboflavin production rate. In addition, it includes $p= 4,088$ covariates, which correspond to the logarithm of the expression levels of $4,088$ bacteria genes. The primary objective of analyzing these data is to identify genes that influence riboflavin production, with the ultimate aim of genetically engineering the bacteria to enhance the yield of riboflavin. Data were collected from $n= 71$ relatively homogeneous samples, which were repeatedly hybridized during a feed batch fermentation process involving different engineered strains and varying fermentation conditions. 

\citet{buhlmann2014high} used stability selection with Lasso and stated that three genes, \texttt{LYSC\_at}, \texttt{YOAB\_at}, and \texttt{YXLD\_at}, have a significant effect on the response variable. \citet{arashi2021ridge} argued that there are some outlier observations in the riboflavin data and proposed a robust version of the ridge regression, known as rank ridge regression, which they applied to the riboflavin data for gene selection. Their analysis identified $41$ relevant genes, including the three genes previously identified by \citet{buhlmann2014high}.

\subsection{Affymetrix Rat Genome 230 2.0 Array}

The second real data is the `Affymetrix Rat Genome 230 2.0 Array' microarray dataset, as introduced by \citet{scheetz2006regulation}. This dataset consists of $n = 120$ twelve-week-old male rats, with expression levels of nearly 32,000 gene probes recorded for each rat. The objective of this analysis is to identify the probes most strongly associated with the expression level of the TRIM32 probe (\texttt{1389163\_at}), which has been linked to the development of Bardet-Biedl syndrome \citep{chiang2006homozygosity}, a genetically heterogeneous disorder that affects multiple organ systems, including the retina. In line with the pre-processing steps described by \citet{huang2008adaptive}, gene probes with a maximum expression level below the $25$th percentile and those with an expression range smaller than $2$ were excluded. This filtering process resulted in a set of $p = 3,083$ gene probes with sufficient expression and variability for further analysis. 

\citet{huang2008adaptive} applied adaptive Lasso \citep{zou2006adaptive} to analyze this dataset and identified $19$ gene probes with non-zero effects on the TRIM32 probe. Later, \citet{li2017variable} demonstrated that four probes—\texttt{1389584\_at}, \texttt{1383996\_at}, \texttt{1382452\_at}, and \texttt{1374106\_at}—are sufficient to model the response variable within a linear framework. With the exception of the third probe, the other probes were also identified by \citet{huang2008adaptive}.

\section{Results}\label{s4}
In this section, we demonstrate the usefulness of Bayesian stability selection using the synthetic and real data introduced in Section \ref{s3}.

\subsection{Synthetic Data}
For the first scenario, we use the elastic net with the mixing parameter $\texttt{alpha}=0.2$. To determine the appropriate $\lambda$ value, we employ a $10$-fold cross-validation using the \texttt{cv.glmnet} function from the \texttt{glmnet} package in R \citep{friedman2010regularization} on the entire dataset $\mathcal{D}$ and determine the regularization value using the 1-se rule. We generate $100$ different datasets based on the setting described in Section \ref{s3} using different random seeds. We set the number of sub-samples $B = 100$ and, at the moment, perform the stability selection without using the Bayesian methodology.

The average of the selection frequencies of the first six variables that are the relevant variables over the $100$ generated datasets $\overline{f_{j}^{\lambda}} = (
  0.529, 0.546, 0.604, 0.609, 0.622, 0.540)^{\top}$. The maximum selection frequency among irrelevant covariates is $0.062$. Although the regression coefficient of the variables in $C_1$ is higher than those in $C_2$, the average selection frequency of the variables in $C_1$ is lower than that of $C_2$. In addition, although the regression coefficient of $x_6$ is more than double that of the variables in $C_2$, its average selection frequency is lower than the variables in $C_2$. This suggests that when using the elastic net, correlated variables, as a group, attract more attention during the selection process compared to uncorrelated variables. In addition, the size of the correlated group directly influences the selection frequencies, with larger groups having higher selection frequencies. 

\citet{meinshausen2010stability} stated that the selection threshold $\pi_{\text{thr}}$ should be selected from the interval $[0.6,0.9]$. This choice is reasonable, as for values near $0.5$, the selection frequency does not exhibit a stable dominance between the selection and non-selection of the variable. In this paper, we adopt a threshold of $0.6$, which represents the most conservative value. Therefore, considering $\pi_\text{thr} = 0.6$, only variables from $C_2$ can be considered stable variables. This underscores the importance of incorporating domain knowledge, which can potentially enhance selection frequencies and help mitigate the ambiguity of the data through prior knowledge and expertise.

Assume that for the first six variables, we have selection frequencies, similar to the average selection frequencies mentioned above, that is, $f_{j}^{\lambda} = \overline{f_{j}^{\lambda}}$. Thus, we can say that $n_j$ equals 53, 55, 60, 61, 62, and 54 for the first six variables. As discussed in Sections \ref{s1} and \ref{s2}, Bayesian stability selection provides a method for involving data owners' knowledge in the analysis in a coherent manner. In this example, if we are informed that for these six variables, the domain expert wants half of the final results to be influenced by background knowledge, that is, $\Tilde{\zeta_j} = 50\%$, and these variables are perceived to have a relevance of $70\%$, that is, $\Tilde{\xi_j} = 70\%$, using Equations~\eqref{eqn:Q1} and~\eqref{eqn:Q2}, we derive $\alpha_j = 70$ and $\beta_j = 30$ for $j \in \{1,2,\dots 6\}$. Consequently, the mean of the posterior selection probabilities for the first six variables $\mathbb{E}(\Pi_{j}^{\lambda} \mid n_{j}, K)$ is 0.615, 0.625, 0.650, 0.655, 0.660, and 0.620. Therefore, based on the posterior means and adopting $\pi_\text{thr} = 0.6$, all six relevant variables can be retrieved.

\begin{figure}[h]
    \centering
    \subfloat[Relevant variables]
{\includegraphics[width=0.5\textwidth]{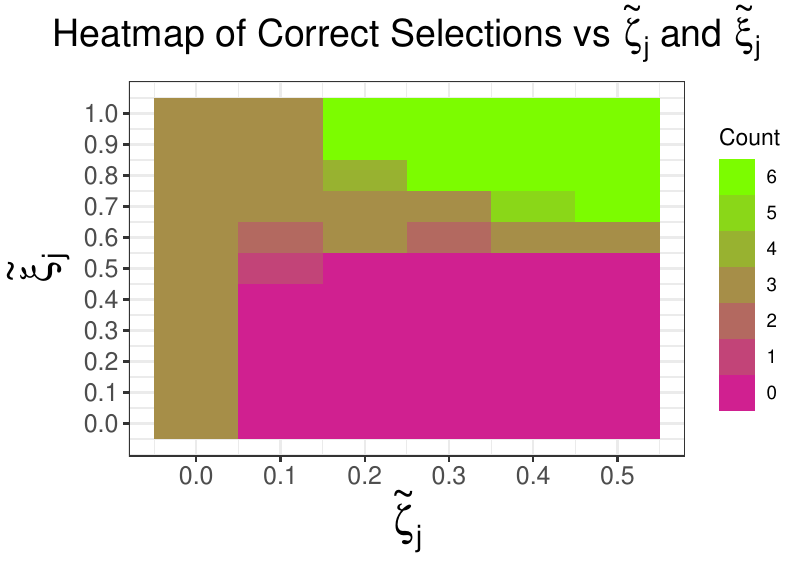}\label{fig1a}}\hfill
    \subfloat[Irrelevant variables]
{\includegraphics[width=0.5\textwidth]{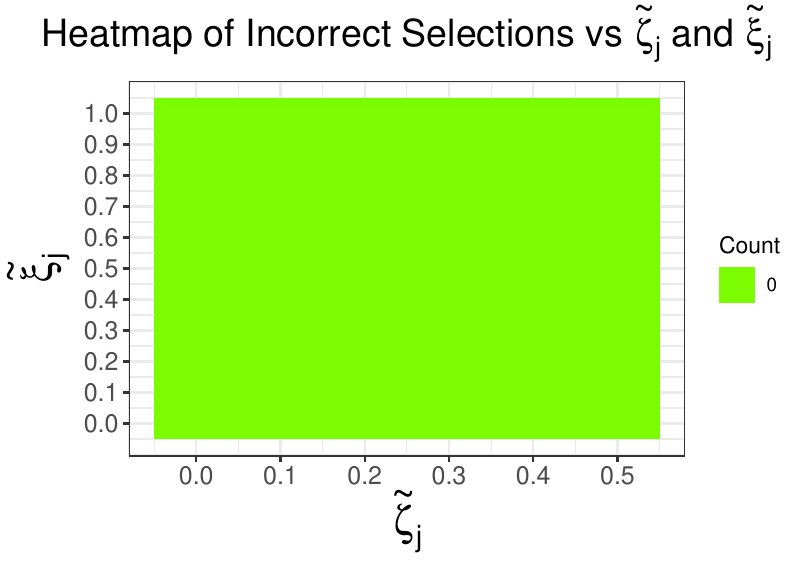}\label{fig1b}}
    \caption{Heatmap of correct and incorrect selections versus $\Tilde{\zeta}_{j}$ and $\Tilde{\xi}_{j}$ in the first scenario}
    \label{fig1}
\end{figure}

Figure \ref{fig1} presents the heat map that depicts correct and incorrect selections in the first scenario. Figure \ref{fig1a} focuses specifically on the first six variables, where it is assumed that the expert's responses to the two proposed questions are the same for these variables. This figure shows the number of variables correctly selected under various combinations of $\Tilde{\zeta}_{j}$ and $\Tilde{\xi}_{j}$. As demonstrated, the optimal outcome occurs when strong prior information regarding the selection of relevant variables is available.

Figure \ref{fig1b} examines the irrelevant variables, totaling 494. Again, it is assumed that the expert's responses to the two questions are the same for all these variables. The figure shows that in all combinations of $\Tilde{\zeta}_{j}$ and $\Tilde{\xi}_{j}$, incorrect prior information suggesting the selection of irrelevant variables does not affect the final decision of the method. This favorable outcome arises from design of the method, which ensures that prior information and data-driven results do not disproportionately influence the final selection decisions.

Following a similar approach to the first scenario, and assuming $f_{j}^{\lambda} = \overline{f_{j}^{\lambda}}$, that is, considering the average of selection frequencies for the first four variables across $100$ datasets as the obtained selection frequencies, the selection frequencies for the second scenario are computed as $0.544$, $0.563$, $0.527$, and $0.43$. This implies that none of these variables can be selected. The largest selection frequency among the irrelevant variables is $0.301$. However, the expected posterior selection probabilities $\mathbb{E}(\Pi_{j}^{\lambda} \mid n_{j}, K)$ for the first four variables, assuming $\Tilde{\zeta} = 50\%$ and $\Tilde{\xi} = 70\%$, are $0.62$, $0.63$, $0.615$, and $0.565$. Consequently, the first three variables can be retrieved.

\begin{figure}[H]
    \centering
    \subfloat[Relevant variables]
{\includegraphics[width=0.5\textwidth]{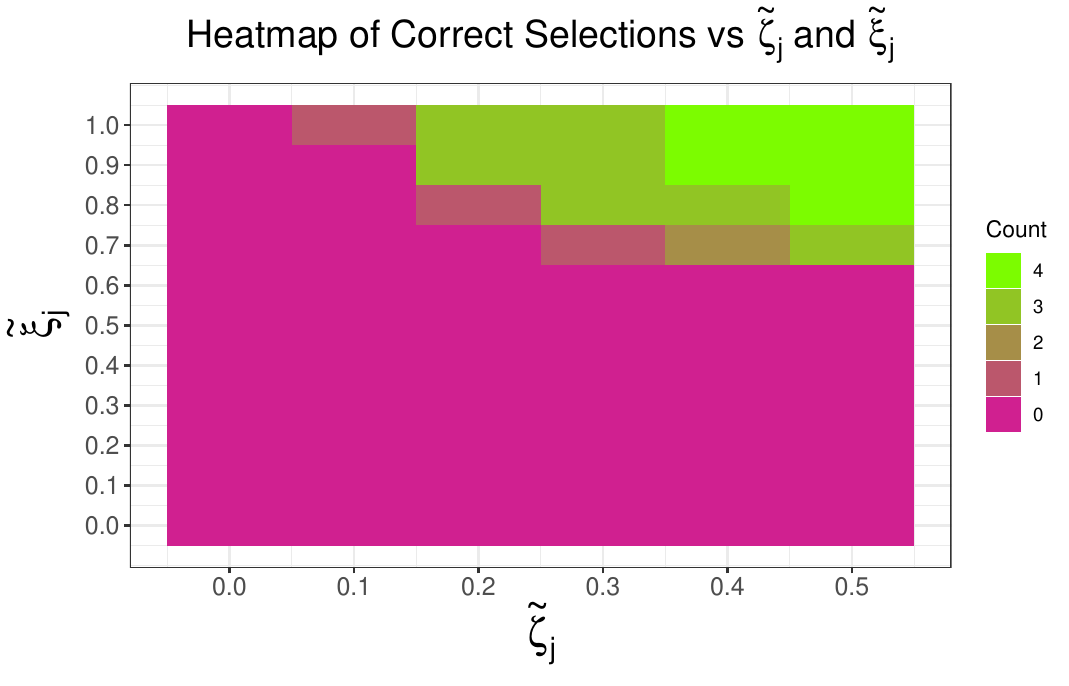}\label{fig2a}}\hfill
    \subfloat[Irrelevant variables]
{\includegraphics[width=0.5\textwidth]{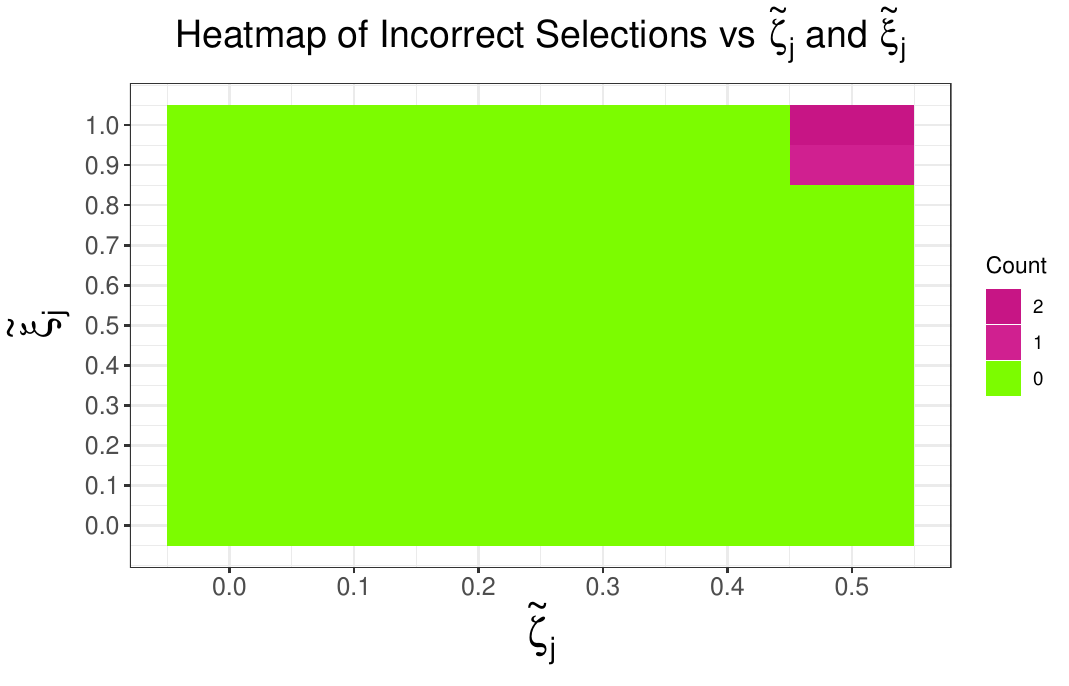}\label{fig2b}}
    \caption{Heatmap of correct and incorrect selections versus $\Tilde{\zeta}_{j}$ and $\Tilde{\xi}_{j}$ in the second scenario}
    \label{fig2}
\end{figure}

Figure \ref{fig2} presents the heat map that depicts correct and incorrect selections in the second scenario, similar to Figure \ref{fig1}. Figure \ref{fig2a} focuses specifically on the first four variables. As demonstrated, the optimal outcome is achieved when strong prior information about the selection of relevant variables is available.

Figure \ref{fig2b} examines the irrelevant variables, totaling 496. The figure shows that in almost all combinations of $\Tilde{\zeta}_{j}$ and $\Tilde{\xi}_{j}$, incorrect prior information suggesting the selection of irrelevant variables does not affect the final decision of the method. Specifically, in the most extreme case, with $\Tilde{\zeta}_{j} = 50\%$ and $\Tilde{\xi}_{j} = 100\%$, only two irrelevant variables are incorrectly selected.

It should be stated that $\Tilde{\zeta}_j$ and $\Tilde{\xi}_j$ must be carefully calibrated by domain experts, as their appropriate values can vary across different variables under selection depending on the state of knowledge.

\subsection{Riboflavin Data}
As discussed in Section \ref{s3}, Riboflavin dataset comprises $p = 4,088$ genes and the goal is to identify genes that influence riboflavin production.

Since \citet{arashi2021ridge} argued that there are some outliers in riboflavin data, we aim to embed a robust selection method within the stability selection framework to evaluate how a robust selector performs on these data, paying attention to outliers. Specifically, we apply Robust Least Angle Regression Selection \citep[RLARS;][]{khan2007robust}, a robust linear model selection technique available through the \texttt{robustHD} R package \citep{robustHD}. In the first attempt, non-informative priors are used for all genes and the number of stability selection iterations $B = 1,000$. The predictor variables are standardized using the \texttt{scale} function prior to being input into RLARS. Genes with posterior means greater than $0.5$ are presented in Table \ref{t1}. Based on Table \ref{t1},  only the gene \texttt{YCKE\_at} is deemed stable.

\begin{table}[h]
\caption{List of genes with posterior means exceeding $0.5$}\label{t1}%
\begin{tabular}{@{}llll@{}}
\hline
Gene & Posterior Mean  & Lower-Bound ($95\%$ CI) & Upper-Bound ($95\%$ CI)\\
\hline
\texttt{YCKE\_at}    & 0.619 & 0.588  & 0.649 \\
\texttt{YOAB\_at}    & 0.589  & 0.558  & 0.619 \\
\texttt{YXLD\_at}    & 0.525  & 0.494  & 0.556 \\
\texttt{YDAR\_at}    & 0.520  & 0.489  & 0.551 \\
\hline
\end{tabular}
\end{table}

In the subsequent attempt, we prioritize the $41$ genes detected by \citet{arashi2021ridge} as effective genes to incorporate background knowledge from their research. We set $\Tilde{\zeta}_{j} = 50\%$ for these $41$ genes, indicating that this background knowledge have a $50\%$ influence on the final inference of these genes. In addition, we assume $\Tilde{\xi}_{j} = 70\%$ for all of them, which means that their perceived relevance are $70\%$. The hyper-parameters $\Tilde{\zeta}_{j}$ and $\Tilde{\xi}_{j}$ should be fine-tuned by genetic experts for greater precision. Since $B=1,000$, we have $\alpha_j = 700$ and $\beta_j = 300$ for these $41$ genes, while for the remaining genes, we use non-informative priors with $\alpha_j = \beta_j = 1$.

Genes with a posterior mean greater than $0.5$ are presented in Table \ref{t2}. Based on Table \ref{t2}, Bayesian stability selection identified three stable genes using informed priors, which are \texttt{YCKE\_at}, 
\texttt{YOAB\_at}, and 
\texttt{YXLD\_at}. The genes \texttt{YOAB\_at} and \texttt{YXLD\_at} were also identified by \citet{buhlmann2014high}; however, the gene \texttt{LYSC\_at} is replaced by \texttt{YCKE\_at}. We interpret this as evidence that the latter is robust to outlier samples. Furthermore, it can be concluded that not all genes identified by \citet{arashi2021ridge} are stable when subject to data perturbations.

\begin{table}[h]
\caption{List of genes with posterior means exceeding $0.5$}\label{t2}%
\begin{tabular}{@{}llll@{}}
\hline
Gene & Posterior Mean  & Lower-Bound ($95\%$ CI) & Upper-Bound ($95\%$ CI)\\
\hline
\texttt{YCKE\_at}    & 0.659 & 0.639  &  0.680 \\
\texttt{YOAB\_at}    & 0.644  & 0.623  & 0.665 \\
\texttt{YXLD\_at}    & 0.613  & 0.591  & 0.634 \\
\texttt{LYSC\_at}    & 0.559  & 0.538  & 0.581 \\
\texttt{YXLE\_at}    & 0.544  & 0.522  & 0.566 \\
\texttt{YDAR\_at}    & 0.520  & 0.489  & 0.551 \\
\texttt{XHLB\_at}    & 0.506  & 0.484  & 0.528 \\
\hline
\end{tabular}
\end{table}

The Bayesian credible intervals presented in Tables \ref{t1} and \ref{t2} have a clear interpretation; given the data and the prior knowledge, the selection probabilities of the variables fall into the intervals provided with probability of $0.95$.

\subsection{Affymetrix Rat Genome 230 2.0 Array}

We apply Bayesian stability selection to the rat genome microarray data introduced in Section \ref{s3}. \citet{li2017variable} identified an outlier in the dataset and subsequently removed it, focusing their analysis on the remaining data. However, we employ a robust selection method that accounts for the presence of the outlier. Following the approach used in the Riboflavin data, we apply RLARS to identify the key probes across the stability selection iterations. We set the number of stability selection iterations $B = 1,000$, and the probes are standardized using the \texttt{scale} function before being incorporated into RLARS. Initially, we use non-informative priors for all probes.

The probes with posterior means greater than $0.3$ are presented in Table \ref{t3}. In particular, none of the four probes listed in Table \ref{t3} were identified by \citet{li2017variable} or \citet{huang2008adaptive}. We interpret this as evidence of the instability of their results when subjected to data perturbation. With a threshold of $\pi_{\text{thr}} = 0.6$, no probe can be considered stable. Among the four probes detected by \citet{li2017variable}, the highest selection frequency corresponds to the probe \texttt{1389584\_at}, with a value of 0.162. Even with $\Tilde{\zeta_j} = 50\%$ and $\Tilde{\xi_j} = 100\%$, this probe has no chance of being classified as stable.
\begin{table}[h]
\caption{List of probes with posterior means exceeding $0.3$}\label{t3}%
\begin{tabular}{@{}llll@{}}
\hline
Probe & Posterior Mean  & Lower-Bound ($95\%$ CI) & Upper-Bound ($95\%$ CI)\\
\hline
\texttt{1389457\_at}    & 0.528   & 0.497  & 0.559  \\
\texttt{1376747\_at}    & 0.523   & 0.492  & 0.554  \\
\texttt{1372928\_at}    & 0.399   & 0.369  & 0.430  \\
\texttt{1390539\_at}    & 0.345   & 0.316  & 0.375  \\
\hline
\end{tabular}
\end{table}

In another approach, we apply the isolation forest model from the \texttt{isotree} R package \citep{isotree} to detect outlier samples. The samples \texttt{GSM130552}, \texttt{GSM130579}, \texttt{GSM130600}, \texttt{GSM130626}, \texttt{GSM130643}, and \texttt{GSM130651} are identified as the 5\% most anomalous observations. Consequently, we omitted these samples in this analysis.

The regularization value $\lambda = 0.0371$ is selected by using a 10-fold cross-validation and the 1-se rule on the entire dataset $\mathcal{D}$. The probes are standardized using the \texttt{scale} function prior to processing by Lasso. Again, we use non-informative priors.

Probes with posterior means greater than $0.2$ are shown in Table \ref{t4}. Among these, only the probe \texttt{1374106\_at} was also identified by \citet{li2017variable} and \citet{huang2008adaptive}. Upon adoption of $\pi_{\text{thr}} = 0.6$, none of the probes is considered stable.
\begin{table}[h]
\caption{List of probes with posterior means exceeding $0.2$}\label{t4}%
\begin{tabular}{@{}llll@{}}
\hline
Probe & Posterior Mean  & Lower-Bound ($95\%$ CI) & Upper-Bound ($95\%$ CI)\\
\hline
\texttt{1390539\_at}    & 0.360   & 0.331  & 0.390  \\
\texttt{1376747\_at}    & 0.218   & 0.193  & 0.244  \\
\texttt{1374106\_at}    & 0.210   & 0.185  & 0.235  \\
\texttt{1393727\_at}    & 0.207   & 0.182  & 0.232  \\
\hline
\end{tabular}
\end{table}

As an example of incorporating prior knowledge, if practitioners have strong beliefs about the importance of probe \texttt{1374106\_at}, based on previous studies such as \citet{huang2008adaptive}, these beliefs can be represented as $\Tilde{\zeta_j} = 50\%$ and $\Tilde{\xi_j} = 100\%$. Under these values, the posterior mean of this probe is $0.605$, indicating that it is stable with the adoption of $\pi_{\text{thr}} = 0.6$. Clearly, the probe \texttt{1390539\_at} requires less strong beliefs about its relevance to be considered stable.

\section{Discussion}\label{s5}

As Christian Hennig noted in the discussion section of \citet{meinshausen2010stability}, it is important in practical applications to incorporate the costs of decisions regarding the inclusion or exclusion of variables into the selection process. A valuable strategy for implementing Bayesian stability selection involves integrating decision costs and conducting risk simulations based on the posterior distributions. By incorporating decision costs, trade-offs can be evaluated between different selection policies and the potential consequences of false positives and false negatives. This approach facilitates a more informed and customized variable selection process, allowing decision makers to simulate various risk scenarios and adjust their selection criteria accordingly. 

In case some groups of functionally related variables are of interest, a prior can be imposed on the probability of including at least one member of a group. This approach facilitates the incorporation of background knowledge about a group into the analysis. The corresponding posterior distribution can be derived in a manner analogous to that used for individual variables, offering valuable insight into the selection behavior of a group as a whole.

Thompson sampling \citep{thompson1933likelihood}, a well-known Bayesian reinforcement learning algorithm, excels in minimizing regret as an agent learns the reward distributions of, for example, a slot machine by sequentially pulling different arms, which is known as the multi-armed bandit problem. Each time the agent selects one or more arms, it updates its beliefs about the reward distributions based on the received rewards, modeled using the Beta distribution. By conceptualizing the binary values of the rows of $M(\lambda)$ as the rewards associated with the selection of variables in each iteration, Bayesian stability selection can be seen as training an expert-informed Thompson-sampling-based agent, where the agent gathers rewards from all arms (that is, variables) at each iteration. Unlike Thompson sampling, where the agent refrains from pulling all arms in every iteration due to associated costs, in the context of Bayesian stability selection, these costs are already accounted for by stability selection, allowing the agent to evaluate all variables simultaneously without such constraints.

One limitation of this paper is the assumption of independence in the selection of variables. This assumption allows us to model the selection status of each variable using an independent Bernoulli distribution. However, a potential improvement would be to account for the interdependencies in variable selection within the modeling process.

\section{Conclusion}\label{s6}
In this paper, we introduced a Bayesian framework for stability selection that integrates expert knowledge with data-driven insights to provide inference on selection probabilities. We specifically argued that selection frequencies should not be interpreted as selection probabilities and proposed a procedure to enhance selection frequencies into inferential selection probabilities. In this context, Bayesian credible intervals are used to quantify the uncertainty regarding the selection status of variables. 

The conjugacy of the proposed distributions removes the need for numerical estimation of posterior distributions, making Bayesian stability selection an efficient aggregation process following the stability selection iterations. We also proposed a two-step process for communication with experts, enabling the elicitation of prior distributions based on their knowledge while also allowing them to control the degree to which their opinions influence the final results. 

Bayesian stability selection remains compatible with existing versions of stability selection, as it is fundamentally built upon the selection matrix, a key component of the original concept introduced by \citet{meinshausen2010stability}. As such, it retains the generality of the stability selection framework and can be applied to a wide range of structure estimation tasks, including variable selection.

\section*{Competing interests}
The authors declare that they have no known competing financial interests or personal relationships that could have appeared to influence the work reported in this paper.

\section*{Authors contributions statement}
 
Mahdi Nouraie was responsible for drafting the manuscript, the development of the research methodology and for writing the computer code used throughout. Samuel Muller and Connor Smith provided critical feedback on the content of the manuscript, refining the clarity and scope of the manuscript and the computer code.  

\section*{Data Availability}
The riboflavin dataset is accessible via the \texttt{hdi} package in R \citep{hdi-package}. The rat microarray data can be obtained from the National Center for Biotechnology Information (NCBI) website at \url{www.ncbi.nlm.nih.gov}, under accession number GSE5680.

The source code used for the paper is accessible through the following GitHub repository: \url{https://github.com/MahdiNouraie/Bayesian-Stability-Selection}.

\section*{Funding}
Mahdi Nouraie was supported by the Macquarie University Research Excellence Scholarship (20213605). Samuel Muller was supported by the Australian Research Council Discovery Project Grant (DP230101908).



\bibliographystyle{abbrvnat}
\bibliography{citation}

\begin{thebibliography}{37}
\providecommand{\natexlab}[1]{#1}
\providecommand{\url}[1]{\texttt{#1}}
\expandafter\ifx\csname urlstyle\endcsname\relax
  \providecommand{\doi}[1]{doi: #1}\else
  \providecommand{\doi}{doi: \begingroup \urlstyle{rm}\Url}\fi

\bibitem[Alfons(2021)]{robustHD}
A.~Alfons.
\newblock {robustHD}: An {R} package for robust regression with high-dimensional data.
\newblock \emph{Journal of Open Source Software}, 6\penalty0 (67):\penalty0 3786, 2021.

\bibitem[Arashi et~al.(2021)Arashi, Roozbeh, Hamzah, and Gasparini]{arashi2021ridge}
M.~Arashi, M.~Roozbeh, N.~A. Hamzah, and M.~Gasparini.
\newblock Ridge regression and its applications in genetic studies.
\newblock \emph{Plos One}, 16\penalty0 (4):\penalty0 e0245376, 2021.

\bibitem[Barbieri and Berger(2004)]{Barbieri2004}
M.~M. Barbieri and J.~O. Berger.
\newblock {Optimal predictive model selection}.
\newblock \emph{The Annals of Statistics}, 32\penalty0 (3):\penalty0 870 -- 897, 2004.

\bibitem[Bayes and Price(1763)]{Bayestheorem}
T.~Bayes and R.~Price.
\newblock Lii. an essay towards solving a problem in the doctrine of chances. by the late rev. mr. bayes, f. r. s. communicated by mr. price, in a letter to john canton, a. m. f. r. s.
\newblock \emph{Philosophical Transactions of the Royal Society of London}, 53:\penalty0 370--418, 1763.

\bibitem[Beinrucker et~al.(2016)Beinrucker, Dogan, and Blanchard]{beinrucker2016extensions}
A.~Beinrucker, {\"U}.~Dogan, and G.~Blanchard.
\newblock Extensions of stability selection using subsamples of observations and covariates.
\newblock \emph{Statistics and Computing}, 26:\penalty0 1059--1077, 2016.

\bibitem[Bodinier et~al.(2023)Bodinier, Filippi, N{\o}st, Chiquet, and Chadeau-Hyam]{bodinier2023automated}
B.~Bodinier, S.~Filippi, T.~H. N{\o}st, J.~Chiquet, and M.~Chadeau-Hyam.
\newblock Automated calibration for stability selection in penalised regression and graphical models.
\newblock \emph{Journal of the Royal Statistical Society Series C: Applied Statistics}, 72\penalty0 (5):\penalty0 1375--1393, 2023.

\bibitem[Bottolo and Richardson(2010)]{bottolo2010evolutionary}
L.~Bottolo and S.~Richardson.
\newblock Evolutionary stochastic search for bayesian model exploration.
\newblock \emph{Bayesian Analysis}, 5\penalty0 (3):\penalty0 583--618, 2010.

\bibitem[B{\"u}hlmann et~al.(2014)B{\"u}hlmann, Kalisch, and Meier]{buhlmann2014high}
P.~B{\"u}hlmann, M.~Kalisch, and L.~Meier.
\newblock High-dimensional statistics with a view toward applications in biology.
\newblock \emph{Annual Review of Statistics and Its Application}, 1\penalty0 (1):\penalty0 255--278, 2014.

\bibitem[Carlin and Louis(1996)]{Carlinemp1996}
B.~P. Carlin and T.~A. Louis.
\newblock \emph{Bayes and Empirical Bayes Methods for Data Analysis}.
\newblock Chapman \& Hall, New York, 1996.

\bibitem[Castillo and Van Der~Vaart(2012)]{castillo2012needles}
I.~Castillo and A.~Van Der~Vaart.
\newblock Needles and straw in a haystack: Posterior concentration for possibly sparse sequences.
\newblock \emph{The Annals of Statistics}, 40\penalty0 (4):\penalty0 2069--2101, 2012.

\bibitem[Chiang et~al.(2006)Chiang, Beck, Yen, Tayeh, Scheetz, Swiderski, Nishimura, Braun, Kim, Huang, et~al.]{chiang2006homozygosity}
A.~P. Chiang, J.~S. Beck, H.-J. Yen, M.~K. Tayeh, T.~E. Scheetz, R.~E. Swiderski, D.~Y. Nishimura, T.~A. Braun, K.-Y.~A. Kim, J.~Huang, et~al.
\newblock {Homozygosity mapping with SNP arrays identifies TRIM32, an E3 ubiquitin ligase, as a Bardet--Biedl syndrome gene (BBS11)}.
\newblock \emph{Proceedings of the National Academy of Sciences}, 103\penalty0 (16):\penalty0 6287--6292, 2006.

\bibitem[Cortes(2024)]{isotree}
D.~Cortes.
\newblock \emph{isotree: Isolation-Based Outlier Detection}, 2024.
\newblock URL \url{https://CRAN.R-project.org/package=isotree}.
\newblock R package version 0.6.1-1.

\bibitem[Dezeure et~al.(2015)Dezeure, B\"uhlmann, Meier, and Meinshausen]{hdi-package}
R.~Dezeure, P.~B\"uhlmann, L.~Meier, and N.~Meinshausen.
\newblock High-dimensional inference: Confidence intervals, p-values and {R}-software {hdi}.
\newblock \emph{Statistical Science}, 30\penalty0 (4):\penalty0 533--558, 2015.

\bibitem[Friedman et~al.(2010)Friedman, Hastie, and Tibshirani]{friedman2010regularization}
J.~Friedman, T.~Hastie, and R.~Tibshirani.
\newblock Regularization paths for generalized linear models via coordinate descent.
\newblock \emph{Journal of Statistical Software}, 33\penalty0 (1):\penalty0 1, 2010.

\bibitem[Hastie et~al.(2009)Hastie, Tibshirani, and Friedman]{hastie2009elementss}
T.~Hastie, R.~Tibshirani, and J.~H. Friedman.
\newblock \emph{The {E}lements of {S}tatistical {L}earning: {D}ata {M}ining, {I}nference, and {P}rediction}, volume~1.
\newblock Springer, New York, 2009.

\bibitem[Hoerl and Kennard(1970)]{hoerl1970ridge}
A.~E. Hoerl and R.~W. Kennard.
\newblock Ridge regression: Biased estimation for nonorthogonal problems.
\newblock \emph{Technometrics}, 12\penalty0 (1):\penalty0 55--67, 1970.

\bibitem[Hu and Johnson(2009)]{hu2009bayesian}
J.~Hu and V.~E. Johnson.
\newblock Bayesian model selection using test statistics.
\newblock \emph{Journal of the Royal Statistical Society Series B: Statistical Methodology}, 71\penalty0 (1):\penalty0 143--158, 2009.

\bibitem[Huang et~al.(2008)Huang, Ma, and Zhang]{huang2008adaptive}
J.~Huang, S.~Ma, and C.-H. Zhang.
\newblock Adaptive {L}asso for sparse high-dimensional regression models.
\newblock \emph{Statistica Sinica}, 18\penalty0 (4):\penalty0 1603--1618, 2008.

\bibitem[Jaynes(2003)]{Jaynes2003}
E.~T. Jaynes.
\newblock \emph{Probability Theory: The Logic of Science}.
\newblock Cambridge University Press, Cambridge, UK, 2003.

\bibitem[Jeffreys(1934)]{jeffreys1934probability}
H.~Jeffreys.
\newblock Probability and scientific method.
\newblock \emph{Proceedings of the Royal Society of London. Series A}, 146\penalty0 (856):\penalty0 9--16, 1934.

\bibitem[Khan et~al.(2007)Khan, Van~Aelst, and Zamar]{khan2007robust}
J.~A. Khan, S.~Van~Aelst, and R.~H. Zamar.
\newblock Robust linear model selection based on least angle regression.
\newblock \emph{Journal of the American Statistical Association}, 102\penalty0 (480):\penalty0 1289--1299, 2007.

\bibitem[Kissel and Mentch(2024)]{Kissel2024}
N.~Kissel and L.~Mentch.
\newblock Forward stability and model path selection.
\newblock \emph{Statistics and Computing}, 34\penalty0 (82), 2024.

\bibitem[Kohn et~al.(2001)Kohn, Smith, and Chan]{kohn2001nonparametric}
R.~Kohn, M.~Smith, and D.~Chan.
\newblock Nonparametric regression using linear combinations of basis functions.
\newblock \emph{Statistics and Computing}, 11:\penalty0 313--322, 2001.

\bibitem[Ley and Steel(2009)]{ley2009effect}
E.~Ley and M.~F. Steel.
\newblock On the effect of prior assumptions in bayesian model averaging with applications to growth regression.
\newblock \emph{Journal of Applied Econometrics}, 24\penalty0 (4):\penalty0 651--674, 2009.

\bibitem[Li et~al.(2017)Li, Liu, and Lou]{li2017variable}
R.~Li, J.~Liu, and L.~Lou.
\newblock Variable selection via partial correlation.
\newblock \emph{Statistica Sinica}, 27\penalty0 (3):\penalty0 983, 2017.

\bibitem[Liang et~al.(2018)Liang, Li, and Zhou]{liang2018bayesian}
F.~Liang, Q.~Li, and L.~Zhou.
\newblock Bayesian neural networks for selection of drug sensitive genes.
\newblock \emph{Journal of the American Statistical Association}, 113\penalty0 (523):\penalty0 955--972, 2018.

\bibitem[Lindley(1975)]{LindleyFuture}
D.~V. Lindley.
\newblock {The Future of Statistics: A Bayesian 21st Century}.
\newblock \emph{Advances in Applied Probability}, 7:\penalty0 106--115, 1975.

\bibitem[Meinshausen and B{\"u}hlmann(2010)]{meinshausen2010stability}
N.~Meinshausen and P.~B{\"u}hlmann.
\newblock Stability selection.
\newblock \emph{Journal of the Royal Statistical Society Series B: Statistical Methodology}, 72\penalty0 (4):\penalty0 417--473, 2010.

\bibitem[Nogueira et~al.(2018)Nogueira, Sechidis, and Brown]{nogueira2018stability}
S.~Nogueira, K.~Sechidis, and G.~Brown.
\newblock On the stability of feature selection algorithms.
\newblock \emph{Journal of Machine Learning Research}, 18\penalty0 (174):\penalty0 1--54, 2018.

\bibitem[Scheetz et~al.(2006)Scheetz, Kim, Swiderski, Philp, Braun, Knudtson, Dorrance, DiBona, Huang, Casavant, et~al.]{scheetz2006regulation}
T.~E. Scheetz, K.-Y.~A. Kim, R.~E. Swiderski, A.~R. Philp, T.~A. Braun, K.~L. Knudtson, A.~M. Dorrance, G.~F. DiBona, J.~Huang, T.~L. Casavant, et~al.
\newblock Regulation of gene expression in the mammalian eye and its relevance to eye disease.
\newblock \emph{Proceedings of the National Academy of Sciences}, 103\penalty0 (39):\penalty0 14429--14434, 2006.

\bibitem[Shah and Samworth(2013)]{shah2013variable}
R.~D. Shah and R.~J. Samworth.
\newblock Variable selection with error control: another look at stability selection.
\newblock \emph{Journal of the Royal Statistical Society Series B: Statistical Methodology}, 75\penalty0 (1):\penalty0 55--80, 2013.

\bibitem[Staerk et~al.(2024)Staerk, Kateri, and Ntzoufras]{staerk2024metropolized}
C.~Staerk, M.~Kateri, and I.~Ntzoufras.
\newblock A metropolized adaptive subspace algorithm for high-dimensional bayesian variable selection.
\newblock \emph{Bayesian Analysis}, 19\penalty0 (1):\penalty0 261--291, 2024.

\bibitem[Thompson(1933)]{thompson1933likelihood}
W.~R. Thompson.
\newblock On the likelihood that one unknown probability exceeds another in view of the evidence of two samples.
\newblock \emph{Biometrika}, 25\penalty0 (3-4):\penalty0 285--294, 1933.

\bibitem[Tibshirani(1996)]{tibshirani1996regression}
R.~Tibshirani.
\newblock Regression shrinkage and selection via the {L}asso.
\newblock \emph{Journal of the Royal Statistical Society Series B: Statistical Methodology}, 58\penalty0 (1):\penalty0 267--288, 1996.

\bibitem[Tikhonov(1963)]{tikhonov1963solution}
A.~N. Tikhonov.
\newblock Solution of incorrectly formulated problems and the regularization method.
\newblock \emph{Soviet Mathematics Doklady}, 4:\penalty0 1035--1038, 1963.

\bibitem[Zou(2006)]{zou2006adaptive}
H.~Zou.
\newblock The adaptive {L}asso and its oracle properties.
\newblock \emph{Journal of the American Statistical Association}, 101\penalty0 (476):\penalty0 1418--1429, 2006.

\bibitem[Zou and Hastie(2005)]{zou2005regularization}
H.~Zou and T.~Hastie.
\newblock Regularization and variable selection via the elastic net.
\newblock \emph{Journal of the Royal Statistical Society Series B: Statistical Methodology}, 67\penalty0 (2):\penalty0 301--320, 2005.

\end{thebibliography}
\end{document}